\documentclass[12pt,preprint]{aastex}
\usepackage{amsmath}
\usepackage{mathtools}
\usepackage{hyperref}
\usepackage{lipsum}

\def\a4{\hsize 17.0cm \vsize 25.cm}

\newcommand{\derp}[2] { \frac{\partial #1}{\partial #2} }

\defcitealias{MRN1977}{MRN} 
\shorttitle{}
\begin{document}

\title{Infrared Observational Manifestations of Young Dusty Super Star Clusters}

\author{Sergio Mart\'inez-Gonz\'alez\altaffilmark{1}, Guillermo Tenorio-Tagle\altaffilmark{1}, Sergiy Silich\altaffilmark{1}}

\altaffiltext{1}{Instituto Nacional de Astrof\'isica \'Optica y
Electr\'onica, AP 51, 72000 Puebla, M\'exico; sergiomtz@inaoep.mx}

\begin{abstract}
The growing evidence pointing at core-collapse supernovae as large dust producers 
makes young massive stellar clusters ideal laboratories to study the evolution of dust immersed into a hot plasma. 
Here we address the stochastic injection of dust by supernovae and follow its evolution due to thermal sputtering 
within the hot and dense plasma generated by young stellar clusters.  
Under these considerations, dust grains are heated by means of random 
collisions with gas particles which results on the appearance of infrared spectral signatures. We present 
time-dependent infrared spectral energy distributions which are to be expected from young stellar clusters. Our 
results are based on hydrodynamic calculations that account for the stochastic injection of dust by supernovae.
These also consider gas and dust radiative cooling, stochastic dust temperature fluctuations, 
the exit of dust grains out of the cluster volume due to the cluster wind and a time-dependent grain size distribution.

\end{abstract}

          
\section{Introduction}
\label{sec:1}

The idea of core-collapse supernovae as major dust producers was first envisaged in the pioneering 
work of \citet*{Cernuschietal1967}. They showed that the effective condensation of refractory 
elements due to the large variation of temperature in the ejecta of core-collapse supernovae can
lead to the formation of massive quantities of dust. However, it took more than two decades until 
SN1987A provided the first direct evidence for the condensation of iron into dust grains 
\citep[][and references therein]{Moseleyetal1989,SuntzeffandBouchet1990,Woodenetal1993,Bautistaetal1995} 
in the SN ejecta. According to \citet{TodiniandFerrara2001} and \citet{Nozawaetal2003}, one can 
expect the formation of ($0.1-1$) M$_{\odot}$ of dust in the first decades after a type II SN event while a 
dust mass fraction between $0.2$-$1.0$ would be destroyed by the supernova reverse shock before being injected 
into the ISM \citep{Nozawaetal2007}. Also, the dust composition consists mostly of silicates and carbon 
(see conflicting interpretations of which composition is dominant by \citet{Matsuuraetal2015} and 
\citet{DwekandArendt2015}). 

These predictions find strong support in recent Herschel and ALMA 
observations of nearby supernova remnants like the Crab Nebula, Cassiopeia A and SN1987A. 
\citet{Gomezetal2012} found evidence for the presence of $0.1-0.25$ M$_{\odot}$ of ejected dust in the 
Crab Nebula, a value that is orders of magnitude larger than what was obtained with Spitzer data 
\citep{Temimetal2012}. \citet{Barlowetal2010} estimated $0.075$ M$_{\odot}$ of cool dust ($\sim 35$ K) 
in the ejecta of Cassiopeia A, however, due to high cirrus contamination along the line of sight, 
they were not able to identify the presence of cold dust ($\sim 20$ K) which 
can increase the content of dust in the ejecta to values in the range of $0.5-1.0$ M$_{\odot}$ 
\citep{Gomez2013}. More recently, \citet{Arendtetal2014} estimated the total mass of dust in the shocked ISM 
and ejecta regions of Cassiopeia A to be $0.04$ M$_{\odot}$, and $\lesssim 0.1$ M$_{\odot}$ in the 
unshocked ejecta.

\citet{Indebetouwetal2014} and \citet{Matsuuraetal2014} fitted the spectral 
energy distribution of SN1987A and derived $\sim 0.8$ M$_{\odot}$ of newly formed dust in the 
ejecta of the supernova with $\sim 0.3$ M$_{\odot}$ of amorphous carbon and $\sim 0.5$ M$_{\odot}$ 
of silicates. On the other hand, \citet{DwekandArendt2015} derived a total mass of dust after day 8500 after 
the explosion of SN1987A consisting of $\sim 0.4$ M$_{\odot}$ of silicates and $\sim 0.05$ M$_{\odot}$ of 
amorphous carbon. 

The large SN rate expected in super star clusters (SSCs) (a few thousand SN events during 
the type II SN era for a $10^5$ M$_{\odot}$ star cluster), together with the large production and injection 
of dust, implies a frequent replenishment of dust inside the star cluster volume \citep{TenorioTagleetal2013}. 
In such clusters, the thermalization of the matter reinserted by massive stars and SNe inside young and massive 
star clusters leads to a large central overpressure and the launching of hot ($\sim 10^{7}$ K) and dense 
($\sim (1-1000)$ cm$^{-3}$) star cluster winds \citep{ChevalierClegg1985,TenorioTagleetal2007}.

These considerations make super star clusters ideal places to heat newly injected dust grains due to 
the transfer of thermal energy from the gas via stochastic collisions with electrons and nuclei as 
discussed by \citet{Dwek1986}. 

Dust grains then cool down in a short time scale and re-emit the obtained energy in the infrared 
regime. This is a very effective cooling mechanism for the hot and dusty gas which can surpass the 
cooling from a gas in collisional ionization equilibrium by several orders of magnitude 
\citep{OstrikerSilk1973,Dwekwerner1981,Dwek1987,Smithetal1996,Guillardetal2009}. 
Infrared excesses have been observed in a considerable number of star clusters in 
low-metallicity blue compact dwarf galaxies, e.g: SBS 0335-052E, Haro 11, Mrk 930 and I Zw18 
\citep{Vanzietal2000,Adamoetal2010, Adamoetal2010b,Adamoetal2011,Fisheretal2014,Izotovetal2014}. 
From these studies, \citet{Vanzietal2000}; \citet{Reinesetal2008} and \citet{Izotovetal2014}, have invoked 
a hot dust component ($\sim 800$ K) in order to explain the near-infrared spectral energy distributions 
observed in the bright SSCs 1 and 2 in SBS 0335-052. 

Here we combine Dwek's (\citeyear{Dwek1986,Dwek1987}) stochastic dust heating and cooling prescriptions with our steady-state wind 
hydrodynamic model to propose several scenarios of dust injection and its influence on the 
spectral energy distributions (SEDs) from starburst regions. To follow the evolution of the dust size
distribution, and therefore the evolution of the SEDs, it is crucial to notice that the stellar winds are steady 
but the rate of supernova makes the dust injection an stochastic process. Dust is injected by stochastic 
supernova events into the intracluster medium, collisionally heated and eroded before the next injection episode.
Moreover, we consider the exit of dust grains as they stream out, coupled to the gas, from the starburst region.

The paper is organized as follows: in Section \ref{sec:2} we describe our star cluster, star cluster wind and 
time-dependent dust size distribution models, and formulate our assumptions regarding the dust grain physics and 
composition. In Section \ref{sec:3}, we use the hydrodynamic results together with the physics of stochastic dust 
heating and cooling to obtain the expected spectral energy distributions of young stellar clusters. In Section 
\ref{sec:4} we summarize our results and outline our conclusions. Complementary information about the dust cooling 
model and stochastic dust temperature fluctuations are presented in Appendix \ref{app:1} and Appendix \ref{app:2}. 

\section{Star Cluster Winds and Dust Injection}
\label{sec:2}
We consider young and massive star clusters in which massive stars follow a generalized Schuster 
stellar density distribution, $\rho_* \propto  [1+(r/R_{c})^2]^{-\beta}$, 
\citep{Palousetal2013,TenorioTagleetal2013,TenorioTagleetal2015}
with $\beta = 1.5$, where $r$ is the distance from the cluster center and $R_{c}$ is the core radius of 
the stellar distribution. This stellar distribution is truncated at radius $R_{SC}$, the star cluster surface. 
Both, $R_{c}$ and $R_{SC}$, define the degree of compactness of the star cluster, which can be measured by 
the radius at which half of the star cluster mass is located, $R_{hm}$. Similarly to \citet{TenorioTagleetal2013}, 
we consider star clusters in which other input parameters are: the star cluster mechanical luminosity, $L_{SC}$,
and the adiabatic wind terminal speed $V_{A\infty}$, which are related to the mass deposition rate 
$\dot{M} = 2 L_{SC}/V_{A\infty}^2$. We assume that the mechanical luminosity scales with 
the total mass of the star cluster, $M_{SC}$, as $L_{SC}= 3 \times 10^{39} (M_{SC}/10^5 \mbox{M}_\odot)$ erg s$^{-1}$ \citep{Leithereretal1999}. 

In our approach, supernova explosions inject dust uniformly throughout the cluster with a
standard \citet[][hereafter MRN]{MRN1977} grain size distribution (dust grain number density in the 
size interval $a$ and $a+\Delta a$):

\begin{eqnarray}
      \label{eq:A1}
      & & \hspace{-1.1cm} 
\derp{n_{i}^{inj}}{a} = A_{i}^{(m)} a^{-\alpha} ,\,\,  a_{min} \leq a \leq   a_{max} ,
\end{eqnarray}

where $a_{min}$ and $a_{max}$ are the minimum grain size and cut-off value of the size distribution. In this 
definition, subindex $i$ is used to distinguish between dust species, in our case graphite and silicate, 
and subindex $m$ numbers the consecutive dust injection events. 

\begin{table}[htp]
\caption{\label{tab:1} Dust Properties}
\begin{tabular}{c c c c }
\hline\hline
Symbol            &        Silicate  &     Graphite     & Definition  \\ \hline
 $\rho_{gr}$      &       3.3        &       2.26       & Dust grain density (g cm$^{-3}$) $^{(1)}$\\
 $\alpha$         &       3.5        &       3.5        & Power index of the \citetalias{MRN1977} distribution  \\
 $f_{i}$          &       0.5        &       0.5        & Mass fraction of the grain species \\
\hline\hline
\end{tabular}\\
$^{(1)}$ \citet{HirashitaandNozawa2013}.
\end{table} 

The normalization factors, $A_{i}^{(m)}$ (with units cm$^{\alpha-4}$), are obtained from the condition:

\begin{eqnarray}
      \label{eq:A2}
A_{i}^{(m)} = \frac{\displaystyle f_{i} M_{dSN}^{(m)}/V_{SC}}{\displaystyle \int^{a_{max}}_{a_{min}} 
     \frac{4}{3} \pi \rho_{gr} a^{3-\alpha} \mbox{ d}a}   ,
\end{eqnarray}

where $\rho_{gr}$ is the dust grain density, $f_{i}$ is the mass fraction of the silicate and graphite 
species, $M_{dSN}^{(m)}$ is the total mass of dust injected in a single supernova event and $V_{SC}$ is 
the star cluster volume. Table \ref{tab:1} summarizes the input parameters for the injected dust size 
distribution and the characteristics of the dust species, in our case, graphite and silicate grains.

The dust lifetime against thermal sputtering at temperatures above 10$^6$ is defined as 
$\tau_{sput}=a/|\dot{a}|$; where $\dot{a}$, the rate at which the dust grain with radius $a$ 
decreases with time $t$ when it is immersed into a hot plasma with temperature $T$ and density 
$n$, is given by \citep{TsaiMathews1995}:

\begin{eqnarray}
      \label{eq:A3}
\dot{a} = \frac{\mbox{ d}a}{\mbox{ d}t} = -1.4 n h \left[\left(\frac{T_{s}}{T}\right)^{w}+1 \right]^{-1}   ,
\end{eqnarray}

which leads to:

\begin{eqnarray}
      \label{eq:A3.1}
\tau_{sput}  =  7.07 \times 10^5 \frac{ a (\mu \text{m})}{n (\text{cm}^{-3})} \left[\left(\frac{T_{s}}{T}\right)^{w}+1 \right]  \text{ yr}  ,
\end{eqnarray}

where $h$, $T_{s}$ and $w$ are constants with values $h=3.2\times 10^{-18}$ cm$^{4}$ s$^{-1}$, 
$T_{s}=2\times 10^6$ K and $w=2.5$. These formulas are an approximation to the detailed calculations of 
\citet{DraineandSalpeter1979} and \citet{Tielensetal1994} for graphite and silicate grains. The continuity
equation which governs the evolution of the dust size distribution due to thermal sputtering is
\citep{LaorDraine1993,YamadaKitayama2005}:

\begin{eqnarray}
\label{eq:A31}
\dot{a}\derp{}{a} \left(\derp{n_{i}}{a} \right) +
\derp{}{t} \left(\derp{n_{i}}{a} \right) 
= 
\begin{cases}   A_{i}^{(m)} a^{-\alpha}/\tau_{inj}^{(m)} 
                , \\  \text{ if } t\leq \tau_{SN}^{(m)}+\tau_{inj}^{(m)} , \\
                0
                , \\  \text{ if } t>\tau_{SN}^{(m)}+\tau_{inj}^{(m)}  ,
\end{cases}    
\end{eqnarray}

where the first case applies for a constant \citetalias{MRN1977} dust injection during a timescale $\tau_{inj}^{(m)}$ 
after the $m$-supernova event has occurred (at $t=\tau_{SN}^{(m)}$); and the second case considers that the $m$-supernova 
dust injection has ceased. The solutions of equations \eqref{eq:A31} after the $n$-supernova event 
(the last event considered) are then:

\begin{eqnarray}
\label{eq:A32}
\derp{n_{i}}{a}
= 
\begin{cases} 
\displaystyle \sum\limits_{m=1}^n \frac{A_{i}^{(m)}}{\tau_{inj}^{(m)}\dot{a}} \left[ \frac{a^{-\alpha+1}}{(-\alpha+1)} 
             - \frac{\left[a-\dot{a}(t-\tau_{SN}^{(m)})\right]^{-\alpha+1}}{(-\alpha+1)} \right] 
             , \\  \text{ if } t\leq \tau_{SN}^{(m)}+\tau_{inj}^{(m)} ,\\
\displaystyle \sum\limits_{m=1}^n  \frac{A_{i}^{(m)}}{\tau_{inj}^{(m)}\dot{a}} \left[ \frac{\left[a-\dot{a}(t-\tau_{SN}^{(m)}-\tau_{inj}^{(m)})\right]^{-\alpha+1}}{(-\alpha+1)}
                - \frac{\left[a-\dot{a}(t-\tau_{SN}^{(m)})\right]^{-\alpha+1}}{(-\alpha+1)} \right] 
             , \\  \text{ if } t>\tau_{SN}^{(m)}+\tau_{inj}^{(m)} , 
\end{cases}    
\end{eqnarray}

with the conditions that $A_{i}^{(m)} = 0$ until the $m$-supernova event occurs and the mass of dust
at $t=\tau_{SN}^{(1)}=0$ equals zero. These general solutions take into account the residual mass of dust 
from the previous injections and the evolved dust size distribution associated to them. Note that the
asymptotic behavior in these solutions is similar to that derived by \citet{Dweketal2008} for the case of
grain destruction with a continuous injection.

\begin{figure}[htbp]
\centering
\epsscale{1.0}
\plotone{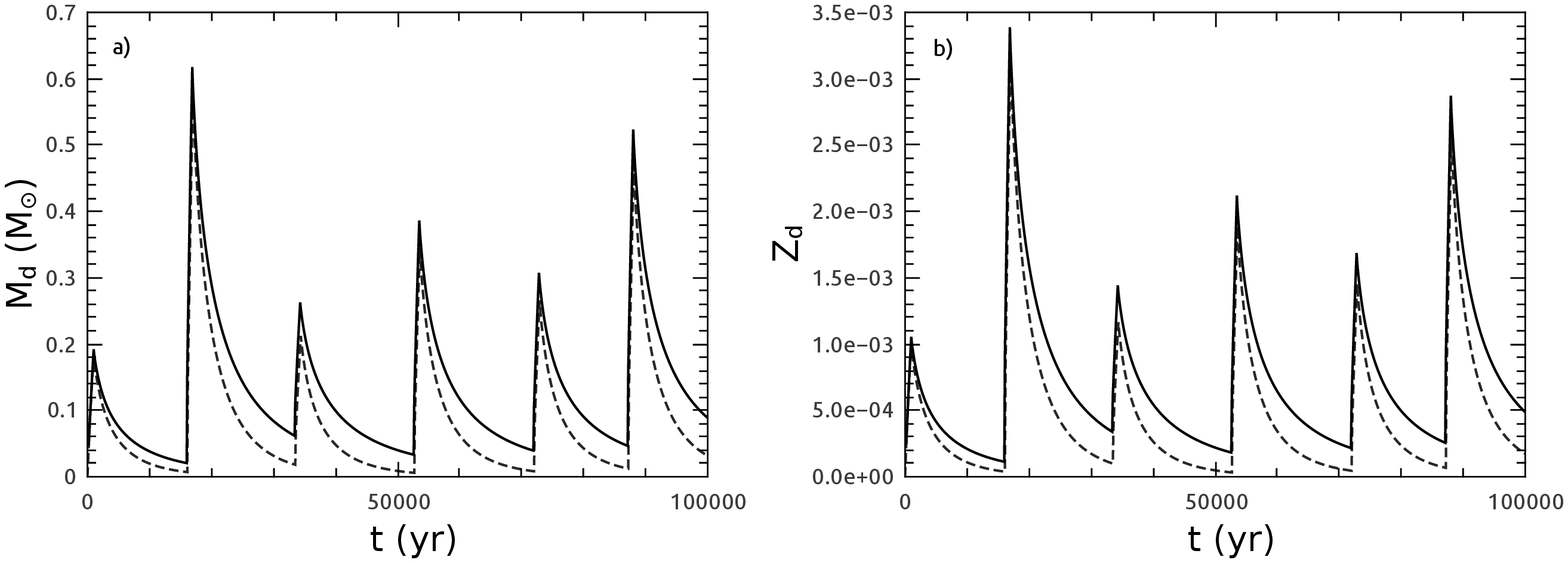}
\caption[Evolution of the dust mass and dust-to-gas mass ratio.]
{Evolution of the dust mass and dust-to-gas mass ratio with and without the exit of dust grains  
from the starburst region. Panels a) and b) show the residual $M_{d}(t)$ and 
dust-to-gas mass ratio, respectively, from 6 injection events with account of thermal sputtering. Solid
lines depict the case when dust grains are subject to thermal sputtering; dashed lines consider also
the case when dust is expelled out of the star cluster. This case corresponds to a 
$3 \times 10^5$ M$_{\odot}$ cluster with $V_{\infty A}= 1000$ km s$^{-1}$, $R_{SC}=5$ pc and $R_{c}=4$ pc. 
The values of $M_{dSN}^{(m)}$ and the interval between consecutive supernova events 
$\Delta \tau_{SN}^{(m)}$ were pseudo-randomly selected. Note that the exit of dust grains
in the cluster wind leads to a more rapid depletion of dust.}
\label{fig:Md}
\end{figure} 

The total mass of dust for each dust species as a function of time is then:

\begin{eqnarray}
\label{eq:Md}
M_{d}(t) = \frac{4 \pi}{3} \rho_{gr} V_{SC} \int_{a_{min}}^{a_{max}} a^3 \derp{n_{i}}{a} \mbox{ d}a , 
\end{eqnarray}

which implies a time-dependent dust-to-gas mass ratio given by:

\begin{eqnarray}
\label{eq:Zd}
Z_{d}(t) = \frac{4 \pi}{3} \frac{\rho_{gr}}{\rho} \int_{a_{min}}^{a_{max}} a^3 \derp{n_{i}}{a} \mbox{ d}a , 
\end{eqnarray}

where $\rho=1.4 m_{H} n$ is the gas mass density and $m_{H}$ is the hydrogen mass.

The above equations do not take into account that dust, independent of its size, is expelled out from 
the cluster and thus $A_{i}^{(m)}$ is no longer a constant. The rate at which dust is ejected from 
the cluster is $\dot{M}_d(t)= 4\pi R_{SC}^2 \rho c_{s} Z_{d}(t)$; where $c_{s}$ is the outflow's local sound 
speed obtained from the wind hydrodynamical calculations. 

In order to consider dust outflowing from the cluster, we have taken a finite differences approach described as 
follows: (1) calculate $M_{d}(t)$ with the original value of $A_{i}^{(m)}$ at $t=\tau_{SN}^{(m)}+\Delta t$; 
(2) at the next time-step, $t=\tau_{SN}^{(m)}+2\Delta t$,  subtract $\dot{M}_d(t) \Delta t$ to $M_{d}(t)$ 
and with this mass, replace $A_{i}^{(m)}$ with

\begin{eqnarray}
      \label{eq:A21}
A_{i}^{(m)'} = \frac{\displaystyle f_{i} \left[ M_{d}(t)-\dot{M}_d(t) \Delta t\right]/V_{SC}}{\displaystyle \frac{1}{A_{i}^{(m)}} \int^{a_{max}}_{a_{min}} 
     \frac{4}{3} \pi \rho_{gr} a^{3} \derp{n_{i}}{a}(t) \mbox{ d}a}   ;
\end{eqnarray}

(3) repeat the procedure for every time-step and for the normalization constants associated to each supernova 
dust injection. In our calculations, we have taken $\Delta t=100$ yr and 
$\tau_{inj}^{(m)}=\tau_{inj}=1000$ yr (the same timescale for every dust injection). We tested this method compared to
the analytic solution (equations \ref{eq:A1}-\ref{eq:Zd}), in the case when $\dot{M}_{d}(t)=0$, and both methods agree very well.

We have selected normally-distributed pseudo-random values for $M_{dSN}^{(m)}$ (except for the first supernova, in 
which $M_{dSN}^{(m)}$ was chosen so that $Z_{d}=10^{-3}$ at $\tau_{inj}$) and $\Delta \tau_{SN}=\tau_{SN}^{(m+1)}-\tau_{SN}^{(m)}$, the interval between supernova events, with 
a mean $0.5$ M$_{\odot}$ and standard deviation $0.15$ M$_{\odot}$ for $M_{dSN}^{(m)}$; and a mean 
interval between supernova explosions ($\sim 17000$ yr for a 10$^5$ M$_{\odot}$ cluster, see Figure \ref{fig:Md}) 
obtained from the supernova rate output of Starburst99 synthesis model \citep{Leithereretal1999} with a 
standard Kroupa initial mass function with lower and upper cut-off masses $0.1$ M$_{\odot}$ and $100$ M$_{\odot}$, 
respectively. 
The standard deviation for $\Delta \tau_{SN}$ was taken to be 10$\%$ of the mean value.

These considerations, imply the presence of a time-dependent reservoir of dust grains embedded into 
the high-temperature ($\sim$ 10$^6$-10$^7$ K) thermalized gas inside the star cluster volume. 

We calculate the gas number density and temperature inside the star cluster by making use of 
our hydrodynamical model \citep[thoroughly discussed in][]{Silichetal2011, Palousetal2013}. Our models 
are quasi-adiabatic, however, they include the effects of gas \citep{Raymondetal1976} and dust radiative 
cooling \citep{Dwek1987} (see Appendix \ref{app:1} for the complete description of the dust cooling calculation). 

Once we know the conditions prevailing inside the star cluster volume (i.e. average values for the gas 
density and temperature), we can apply \citet{Dwek1986} prescriptions to calculate the temperature 
distribution, $G(a, T_{d})$, of dust grains which follow different dust size distributions. The infrared 
flux per unit wavelength, produced by a population of dust grains with the same chemical composition, 
from a star cluster located at distance $D_{SC}$, can then be calculated as \citep{DwekandArendt1992}:

\begin{eqnarray}
      \label{flambda}
       \hspace{-1.1cm}
f_{\lambda} &=& \left(\frac{1.4 m_{H} Z_{d} N_{H}}{\rho_{d}} \right) \pi \Omega_{SC} \int_{a_{min}}^{a_{max}}
            \int_{0}^{\infty} a^{2} \derp{n_{i}}{a} Q_{abs}(\lambda,a) B_{\lambda}(T_d) G(a,T_d) \mbox{ d}T_{d} \mbox{ d}a 
\end{eqnarray}

in units erg s$^{-1}$ cm$^{-2}$ \AA$^{-1}$, or alternatively, in units erg s$^{-1}$ cm$^{-2}$ Hz$^{-1}$ or Jansky, if one 
is interested in the flux per unit frequency, $f_{\nu}$, where $\lambda f_{\lambda}=\nu f_{\nu}$. Since both quantities, 
$f_{\lambda}$ and $f_{\nu}$, are widely used by different authors 
\citep[e.g.][]{Reinesetal2008,Adamoetal2010b, Fisheretal2014,Izotovetal2014}, we present them both in all our figures, 
taking into account the contribution from graphite and silicate grains. In the above equation, $N_{H}$ is the hydrogen 
column density through the star cluster volume ($= 4/3 n R_{SC}$), $a$ is the dust grain radius, 
$\rho_{d} = 4/3\pi \rho_{gr} \displaystyle \int_{a_{min}}^{a_{max}} a^{3} \derp{n_{i}}{a} \mbox{ d}a $ is the 
size-averaged dust density, and $\Omega_{SC}=\pi (R_{SC}/D_{SC})^2$, is the solid angle subtended by the star cluster. Additionally, $T_{d}$ is the dust 
temperature, $G(a,T_d)$ is the dust temperature distribution resulting from stochastic temperature fluctuations, 
$Q_{abs}(\lambda,a)$ is the dust absorption efficiency and $B_{\lambda}(T_{d})$ is the Planck function in terms of the 
wavelength, $\lambda$. In our models we have set the distance to the star cluster as $D_{SC}=10$ Mpc. 
A complete discussion of the stochastic dust temperature fluctuations is presented in Appendix \ref{app:2}.

In all our calculations, we neglected the charge of dust grains \citep{Smithetal1996} as well as the 
contribution to the infrared flux from dust grains outside the star cluster volume. We are also not dealing with possible effects related to the absorption of ionizing photons by 
dust grains which could be important as long as the ionizing flux from the star cluster is strong. However, coeval clusters suffer a substantial
reduction of their ionizing photon flux as soon as they enter the SN era. The number of emitted UV photons falls with time, as $t^{-5}$ 
\citep{Beltramettietal1982} and thus after $5-6$ Myr the number of UV photons is almost two orders of magnitude smaller than at the start of 
the evolution. This fall in the ionizing flux reduces the time during which the UV radiation may be more important than gas-dust 
collisions which could be important during all the type II SN era (from $\sim 3$ to $40$ Myr). The
hydrodynamical model assumes that dust grains move with the same velocity as the injected gas and thus
we have not considered the effects of kinetic sputtering in the intracluster medium.


\section{Infrared Spectral Energy Distributions}
\label{sec:3}

To assess the impact that collisional heating of dust grains have on the expected infrared 
spectral energy distributions from young and massive dusty star clusters, we have run 
several models with the input parameters described in the previous section 
($L_{SC}$, $V_{A\infty}$, $R_{c}$, $R_{SC}$, $a_{min}$, $a_{max}$ and $t$). 
Our reference model \textit{A}, consists of a star cluster with a total mass of 
$10^5$ M$_{\odot}$ (which corresponds to a mechanical luminosity $3 \times 10^{39}$ erg s$^{-1}$), 
$R_{hm}=3.92$ pc (obtained from values $R_{c}=4$ pc and $R_{SC}=5$ pc), an adiabatic wind terminal speed 
$V_{A\infty}=1000$ km s$^{-1}$, lower and upper limits for the injected dust size distribution, 
$a_{min}= 0.001 \mu$m and $a_{max}= 0.5 \mu$m, respectively; and an equal mixture of graphite and 
silicate grains. The other models vary one or more of the input parameters with respect to 
model \textit{A}. Models \textit{B}-\textit{C} explore different values of the mechanical luminosity, 
models \textit{D} and \textit{E} vary the adiabatic wind terminal speed and models \textit{F} and 
\textit{G} differ in the compactness of the star cluster. The reference model \textit{A}, 
as well as models \textit{B}-\textit{G} are evaluated at the end of the first injection event 
($t=1000$ yr) where, as pointed out before, $M_{dSN}^{(1)}$ is set to allow $Z_{d}(\tau_{inj})$ to 
be equal to $10^{-3}$. Models \textit{A1}-\textit{A4} are evaluated at later times. Table \ref{tab:2} 
presents the input parameters for our 11 models.

\begin{table}[htp]
\caption{\label{tab:2} Model Parameters}
\begin{tabular}{c c c c c c c c}
\hline\hline
Model      &     t     & $R_{c}$ & $R_{SC}$ &   $R_{hm}$ &    $L_{SC}$       & $V_{A \infty}$ &   $ Z_{d}    $       \\
            &   (yr)    &  (pc)   &   (pc)   &      (pc)  &(erg s$^{-1}$)     &  (km s$^{-1}$) &      ($10^{-3}$)       \\ \hline
\textit{A}  &   1000    &   $4$   &   $5$    &   $3.52$   & $3 \times 10^{39}$ &      1000      & $1.0$       \\
\textit{A1} &  17000    &   $4$   &   $5$    &   $3.52$   & $3 \times 10^{39}$ &      1000      & $3.1$       \\
\textit{A2} &  17500    &   $4$   &   $5$    &   $3.52$   & $3 \times 10^{39}$ &      1000      & $2.5$       \\
\textit{A3} &  25000    &   $4$   &   $5$    &   $3.52$   & $3 \times 10^{39}$ &      1000      & $0.4$       \\
\textit{A4} &  33000    &   $4$   &   $5$    &   $3.52$   & $3 \times 10^{39}$ &      1000      & $0.1$       \\
\textit{B}  &   1000    &   $4$   &   $5$    &   $3.52$   & $1 \times 10^{39}$ &      1000      & $1.0$       \\
\textit{C}  &   1000    &   $4$   &   $5$    &   $3.52$   & $9 \times 10^{39}$ &      1000      & $1.0$       \\
\textit{D}  &   1000    &   $4$   &   $5$    &   $3.52$   & $3 \times 10^{39}$ &       750      & $1.0$       \\
\textit{E}  &   1000    &   $4$   &   $5$    &   $3.52$   & $3 \times 10^{39}$ &      1500      & $1.0$       \\
\textit{F}  &   1000    &   $2$   &   $5$    &   $2.98$   & $3 \times 10^{39}$ &      1000      & $1.0$       \\
\textit{G}  &   1000    &   $4$   &   $7$    &   $4.59$   & $3 \times 10^{39}$ &      1000      & $1.0$       \\
\hline\hline
\end{tabular}
\end{table}

\subsection{The Reference Model}
\label{sec:3.0}

\begin{figure}[htbp]
\centering
\epsscale{1.0}
\plotone{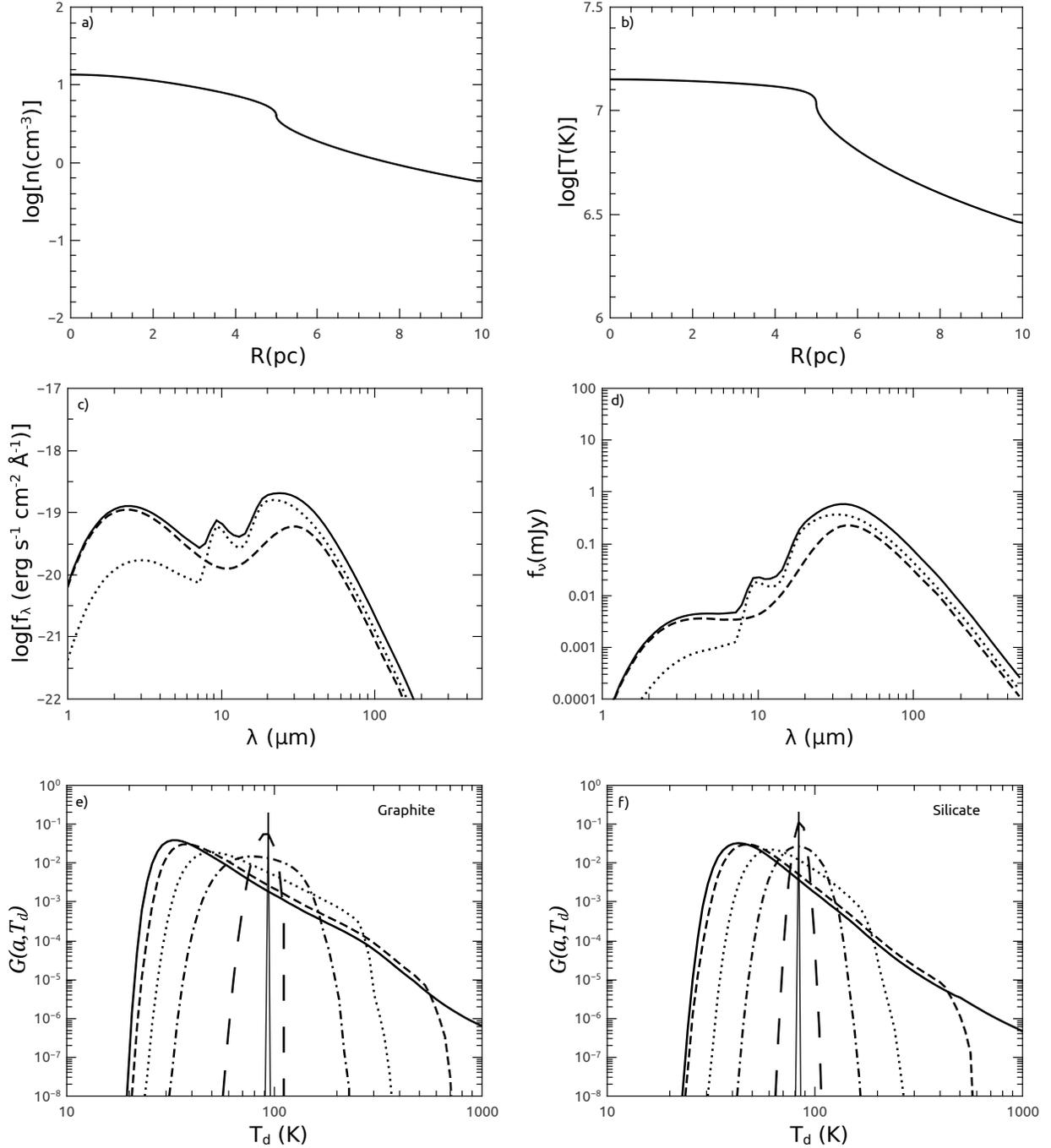}
\caption{The Reference Model. Top panels a) and b), show the gas density and temperature radial profiles for our
reference model \textit{A}. Panels c) and d) present the fluxes per unit wavelength, $f_{\lambda}$, and per
unit frequency, $f_{\nu}$, respectively. The dashed line depicts the contribution from graphite grains, the dotted
line the contribution from silicate grains while the solid line comprises both contributions. Bottom panels e) 
and f), present the dust temperature distribution for different grain sizes for graphite and silicate grains,
respectively. In the bottom panels, the solid, dashed, dotted, dashed-dotted, long-dashed and the delta-like
curves correspond to sizes $0.001$, $0.002$, $0.01$, $0.05$, $0.1$ and $0.5$ $\mu$m, respectively.}
\label{fig:Ref}
\end{figure}

The outcomes from our reference model \textit{A} are displayed in Figure \ref{fig:Ref}. 
In this case, the prevailing conditions inside the star cluster are: an average value for the gas density $\sim 10$ cm$^{-3}$ 
and an average gas temperature $\sim 1.35 \times 10^7$ K. From those conditions, we computed the dust temperature distributions, 
$G(a,T_{d})$, for different dust sizes, and calculated the resultant flux averaged by the size distribution of graphite and silicate 
grains (see Appendix \ref{app:1}). In order to quantify the contribution to the total flux, we display separate fluxes from 
graphite and from silicate grains. As shown in panels e) and f), small grains ($\lesssim 0.05$ $\mu$m) are more likely to undergo
strong temperature fluctuations and therefore span a wide range of temperatures 
(from a few $\sim 10$ K to a few $\sim 1000$ K for grains with $a=0.001$ $\mu$m, making them to strongly emit in 
all near-infrared (NIR), mid-infrared (MIR) and far-infrared (FIR) wavelengths) due to their low heat capacities 
(which scale as $\sim a^3$) and small cross sections. 

Big grains ($\gtrsim 0.1$ $\mu$m) with larger cross sections (which make them subject to more frequent collisions) emit nearly 
as a blackbody at their equilibrium temperature. Intermediate-size grains ($0.05 \gtrsim a \gtrsim 0.1$) exhibit a combination 
of both behaviors. Hence, the emission from $1$ to $8$ $\mu$m is dominated by hot and small graphite grains. Between $9$ and 
$14$ $\mu$m, the emission is dominated by the 10-micron broad feature, associated to the dust absorption efficiency, 
$Q_{abs}(\lambda,a)$, of silicate grains. The emission from $15$ to several hundred microns peaks around $\sim 35$ $\mu$m; 
it is dominated by big grains near their equilibrium temperature ($\sim 93$ K for graphite grains and $\sim 75$ K for 
silicate grains) and by small grains with temperatures ranging from $\sim 10$ K to $\sim 100$ K. Note that the emission from 
graphite and silicate grains is almost identical for $\lambda \gtrsim 35$ $\mu$m because their dust temperature distributions 
are very similar. In this case, the mass of dust inside the star cluster volume is $M_{d}(t=\tau_{inj})=0.19$ M$_{\odot}$.

\begin{figure}[htbp]
 \epsscale{1.0}
 \plotone{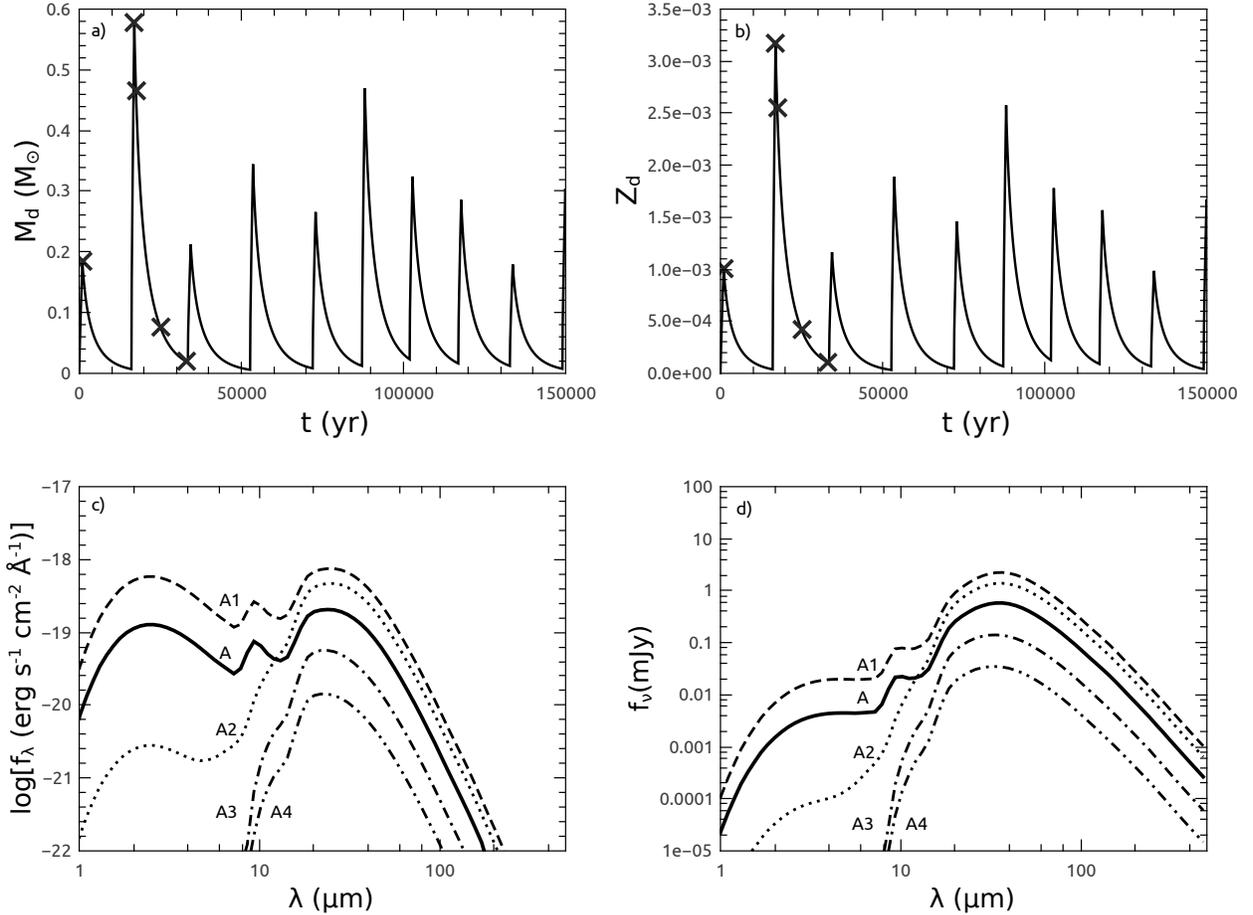}
 \caption[Spectral Energy Distributions for models \textit{A}, \textit{A1}, \textit{A2}, \textit{A3} 
 and \textit{A4}.]{Spectral Energy Distributions for models \textit{A}, \textit{A1}, \textit{A2}, \textit{A3} 
 and \textit{A4}. Top panels a) and b) show the evolution
 of the dust mass and dust-to-gas mass ratio, respectively, during 9 injection events taking into account both dust
 sputtering and their exit out of the cluster as a wind. The times at which models \textit{A1}-\textit{A4} were evaluated
 are marked with crosses. Bottom panels c) and d) present the 
 values of the fluxes per unit wavelength, $f_{\lambda}$, and per unit frequency, $f_{\nu}$, for each 
 evolved models, respectively. Solid, dashed, dotted, dashed-dotted and dashed-double-dotted lines depict the SEDs for
 models \textit{A}, \textit{A1}, \textit{A2}, \textit{A3} and \textit{A4}, respectively. Note that the strong emission present
 during dust injection rapidly vanishes when dust injection has ceased.}
 \label{fig:SEDS}
\end{figure} 

We have evaluated our reference model \textit{A} at four later times (models \textit{A1},
\textit{A2}, \textit{A3} and \textit{A4} evaluated at $\sim 17000$, $17500$, $25000$, and $33000$ yr, 
respectively). When dust injection is not taking place (models \textit{A2}-\textit{A3}), the dust size 
distribution and the dust-to-gas mass ratio rapidly evolve and greatly depart from the injected dust 
size distribution as a consequence of the short timescale for thermal sputtering (see equation \ref{eq:A3.1}). This 
is reflected in a lack of small grains and therefore, the NIR excesses noted in models \textit{A} and 
\textit{A1} (evaluated just before the end of the first and second dust injection episodes, respectively) 
rapidly vanish. This situation is illustrated in Figure \ref{fig:SEDS} which shows evolving spectral energy distributions for 
models \textit{A}, \textit{A1}, \textit{A2}, \textit{A3} and \textit{A4}.
Top panels show the evolution of the dust mass (panel a) and dust-to-gas mass ratio (panel b) during 9 
injection events; the times at which these models were evaluated are marked with crosses. Bottom panels show 
evolving spectral energy distributions at the end of the first dust injection ($1000$ years, model
\textit{A}), at the end of the second dust injection. A strong emission at NIR and MIR wavelengths is 
present during dust injection, however, when dust injection has ceased, this strong emission rapidly 
vanishes which is notorious just $500$ years after dust injection (model \textit{A2}) due to the 
depletion of grains with radius $\lesssim 0.01$ $\mu$m. Models \textit{A3} and \textit{A4}, which are 
more affected by thermal sputtering and the exit of dust grains, have diminished dust-to-gas mass ratios 
and a negligible emission at NIR-MIR wavelengths. 

If one compares flux ratios, e.g. $f_{\nu}(25 \mu \text{m})/f_{\nu}(100 \mu \text{m})$ and 
$f_{\nu}(3.5 \mu \text{m})/f_{\nu}(25 \mu \text{m})$, one can note that the former is almost a constant
around $10$, while the latter rapidly approaches to zero. This is explained due to the fact that the emission 
at $25$ $\mu$m and $100$ $\mu$m originates mainly from big grains radiating at their equilibrium temperature 
(which are less affected by thermal sputtering), while the emission at $3.5$ $\mu$m comes from 
stochastically-heated small grains which are strongly affected by thermal sputtering. The mass of dust in 
models \textit{A1}-\textit{A4} is $0.61$, $0.51$, $0.14$ and $0.06$ M$_{\odot}$, respectively. We note that 
in these models, roughly $20\%$ of the total mass injected by each supernova explosion is expelled out 
from the star cluster as a wind. 

\subsection{Models with Different Star Cluster Mechanical Luminosities}
\label{sec:3.3}

 \begin{figure}[htbp]
 \epsscale{1.0}
 \plotone{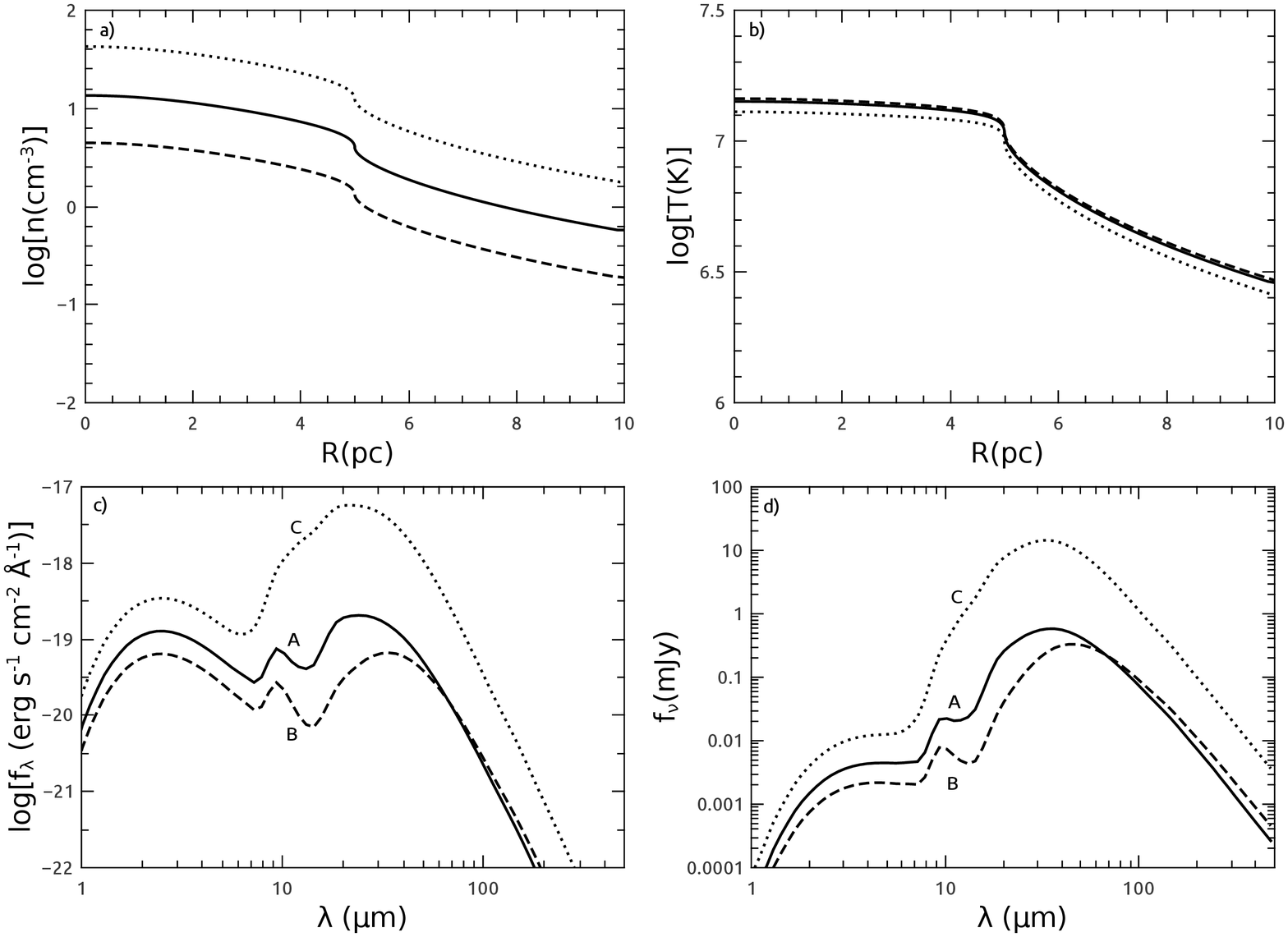}
 \caption[Results for models with different mechanical luminosities.]{Results for models with different mechanical luminosities 
 \textit{A} ($3\times10^{39}$ erg s$^{-1}$,  solid curves), \textit{B} ($1\times10^{39}$ erg s$^{-1}$, dashed curves) and 
 \textit{C} ($9 \times10^{39}$ erg s$^{-1}$, dotted curves). Panels a) and b) show the gas number density  and temperature profiles 
 of each model, respectively. Panels c) and d) present the values of the fluxes per unit wavelength, $f_{\lambda}$, and per unit frequency, 
 $f_{\nu}$, for each model, respectively. Note that an increase in the star cluster mechanical luminosity (or equivalently in the mass
 of the star cluster), leads to an increase in the infrared emission from dust grains.
 }
 \label{fig:Lsc}
\end{figure}

The results derived from models \textit{A},\textit{B} and \textit{C} are shown in Figure \ref{fig:Lsc}. These models
(as well as models \textit{D}-\textit{G}) are evaluated at $t=\tau_{inj}=1000$ yr. Models \textit{B} and \textit{C} consider
three times smaller ($L_{SC} = 1 \times 10^{39}$ erg s$^{-1}$) and three times larger ($L_{SC} = 9 \times 10^{39}$ erg s$^{-1}$) values of the star
cluster mechanical luminosity than what is considered in model \textit{A}, respectively. An increase in the value of $L_{SC}$, 
yields a higher gas density inside the star cluster, which results in more frequent gas-grain collisions leading to an increase in the 
infrared emission (see equations \ref{eq:taucoll} and \ref{eq:Ge} in Appendix \ref{app:2}). As more 
massive clusters are considered, a higher supernova rate is expected and therefore, there is less time between 
supernova episodes to erode dust grains. This leads to a more persistent dust reservoir at all times 
which is reflected in an enhancement of the infrared spectrum. Thus, model \textit{C} surpasses the infrared emission from 
models \textit{A} and \textit{B}. However, model \textit{C} is also more affected by thermal sputtering 
which is noticeable by a decreased emission at $\lambda \lesssim 15$ $\mu$m. The mass of dust inside $R_{SC}$ 
at $t=\tau_{inj}$ is $6.0 \times 10^{-2}$ M$_{\odot}$ and $0.57$ M$_{\odot}$ for models \textit{B} and \textit{C}, 
respectively.

\subsection{Models with Different Adiabatic Terminal Speeds}
\label{sec:3.4}

\begin{figure}[htbp]
 \epsscale{1.0}
 \plotone{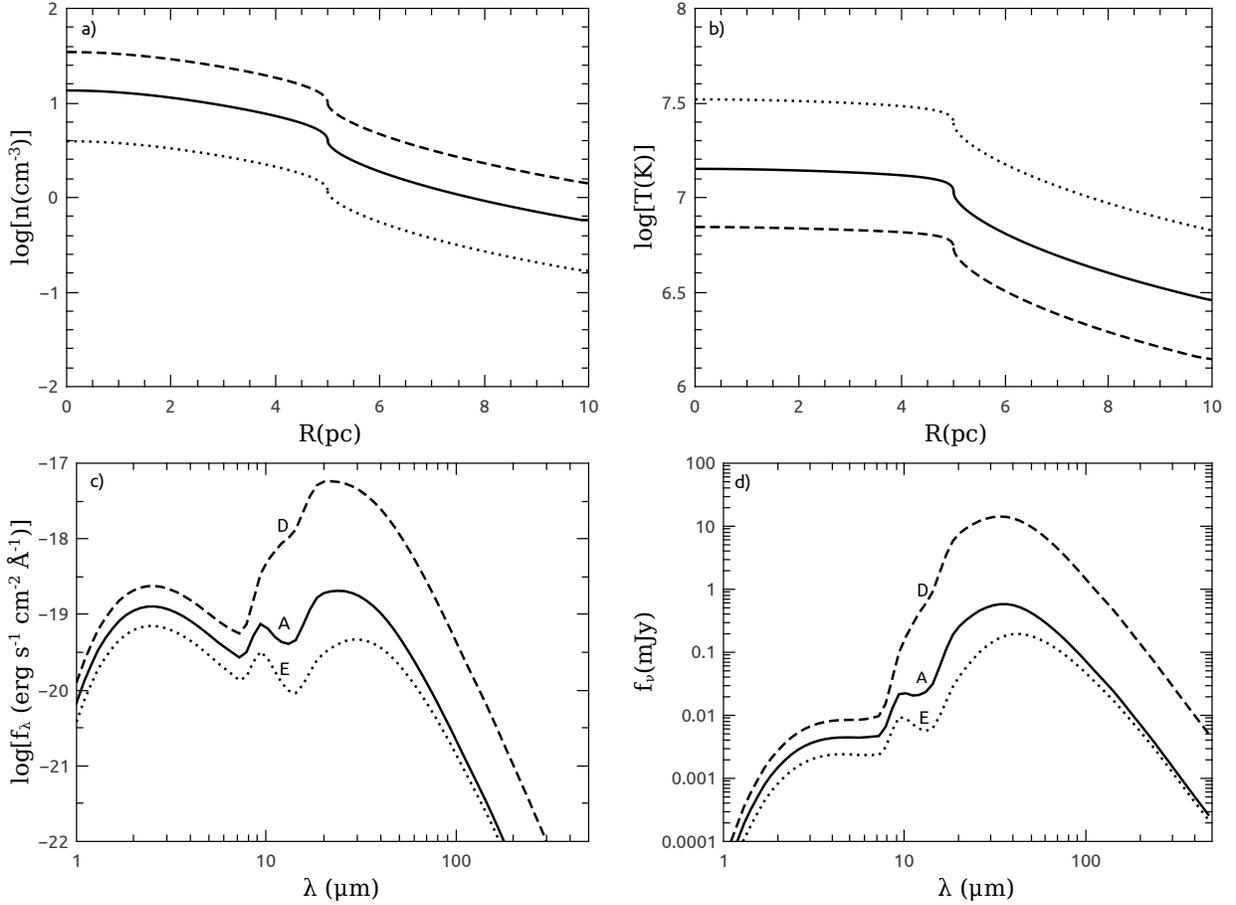}
 \caption[Same as Figure \ref{fig:Lsc} but for models with different adiabatic wind terminal speeds.]
 {Same as Figure \ref{fig:Lsc} but for models with different adiabatic wind terminal speeds \textit{A} 
 ($1000$ km s$^{-1}$,  solid curves), \textit{D} ($750$ km s$^{-1}$, dashed curves) and \textit{E} 
 ($1500$ km s$^{-1}$, dotted curves). Note that an increase  in the adiabatic terminal speed, lead to a decrease 
 in the dust emission inside the star cluster volume.}
 \label{fig:Vterm}
\end{figure}

Figure \ref{fig:Vterm} shows the results obtained from models \textit{A,G} and \textit{E}. In these models we examine different values of the  
adiabatic wind terminal speed. As the gas density decreases with an increasing adiabatic wind terminal speed, the characteristic 
time between successive electron collisions with a dust grain increases (see equation \ref{eq:taucoll} in 
Appendix \ref{app:2}). Dust grains then are less heated and their emission decreases (model \textit{E} compared to model 
\textit{A}). The opposite situation occurs in models with a lower value of the adiabatic wind terminal speed 
(model \textit{D} compared to model \textit{A}) which also causes a decreased in the emission at 
$\lambda \lesssim 15$ $\mu$m provoked by the depletion of small grains by the action of thermal 
sputtering in a denser medium. The mass of dust inside $R_{SC}$ is $0.47$ M$_{\odot}$ and $5.29 \times 10^{-2}$ M$_{\odot}$ for models 
\textit{D} and \textit{E}, respectively.

\subsection{Models for Different Cluster Sizes}
\label{sec:3.5}

\begin{figure}[htbp]
 \epsscale{1.0}
 \plotone{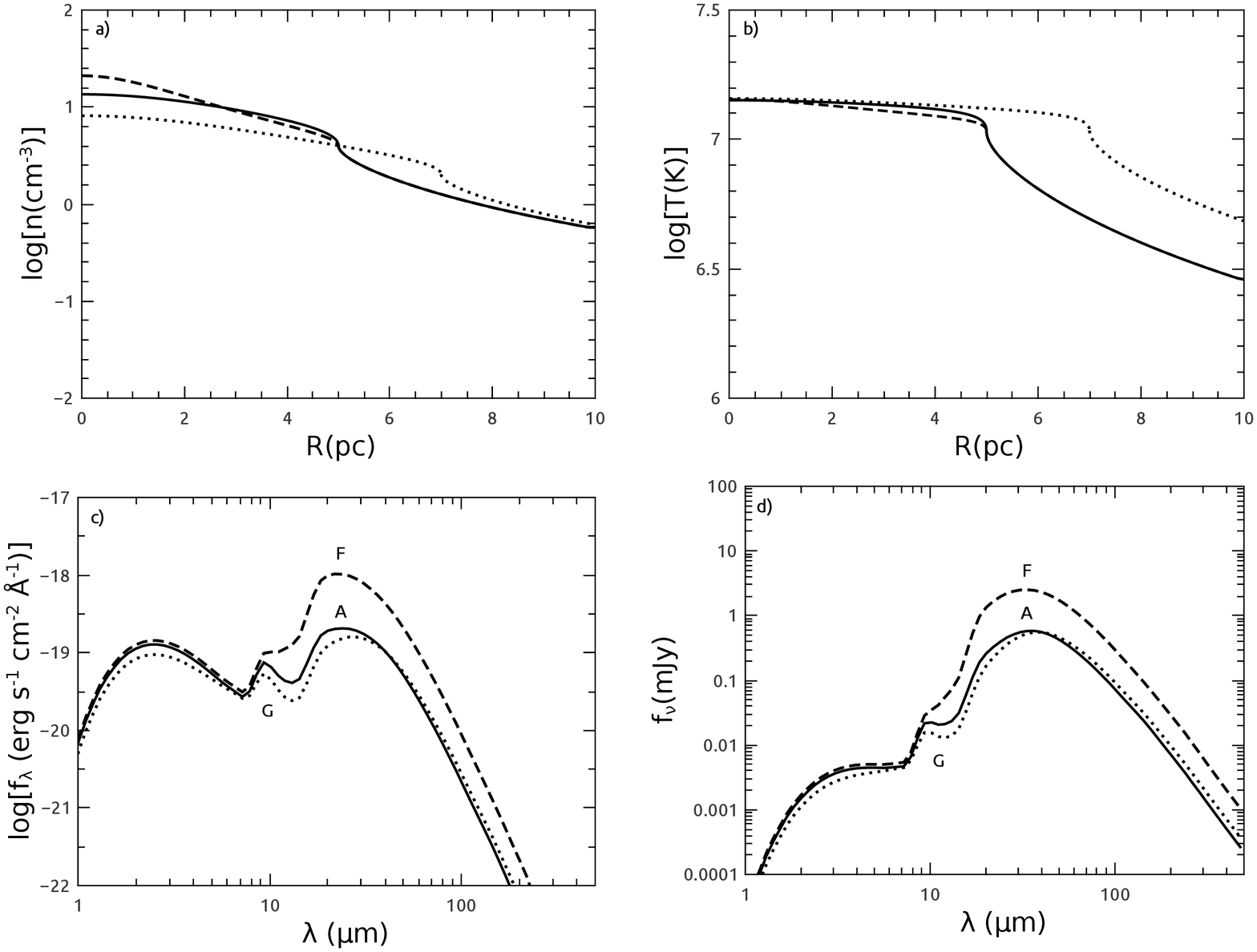}
 \caption[Same as Figure \ref{fig:Lsc} but for models with different star cluster sizes.]
 {Same as Figure \ref{fig:Lsc} but for models with different star cluster sizes. Solid lines correspond to model 
 \textit{A} ($R_{c}=4$ pc, $R_{SC}=5$ pc and $R_{hm}=3.92$ pc), dashed lines display the results obtained from model \textit{F} 
 ($R_{c}=2$ pc, $R_{SC}=5$ pc and $R_{hm}=3.52$ pc), and dotted lines show the results from model \textit{G} 
 ($R_{c}=4$ pc, $R_{SC}=7$ pc and $R_{hm}=5.45$ pc). Note that more compact clusters lead to an increase in 
 the dust emission inside the star cluster volume, especially at FIR wavelengths.}
 \label{fig:Rsc}
\end{figure}

We now focus on models with different values of $R_{c}$ and $R_{SC}$ (see Figure 
\ref{fig:Rsc}). Model \textit{F} is evaluated with a smaller star cluster core 
radius $R_{c} = 2$ pc. Model \textit{G} corresponds to a star cluster with a larger 
cut-off radius $R_{SC} = 7$ pc. A more compact cluster, as in model \textit{F} 
compared to model \textit{A}, yields an enhanced flux at all 
wavelengths, however, this effect is more noticeable at FIR wavelengths where the role
of thermal sputtering is less important. The situation is different when one considers a 
less compact cluster (model \textit{G} compared to model \textit{A}), when the gas number 
density is decreased and therefore, the dust emission is diminished. In these models, half of 
the star cluster mass is located inside $3.52$ pc, $2.98$ pc and $4.59$ pc for 
models \textit{A}, \textit{F} and \textit{G}, respectively. The mass of dust inside 
the star cluster volume is $0.21$ M$_{\odot}$ and $8.62 \times 10^{-2}$ M$_{\odot}$ 
for models \textit{F} and \textit{G}, respectively.

\section{Conclusions}
\label{sec:4}

Motivated by the abundant evidence for core-collapse supernovae as major dust producers, and the large 
SN rate expected in young massive star clusters, we have studied the frequent injection of dust 
grains into the plasma interior of super star clusters, which become ideal places to heat dust 
grains by means of random gas-grain collisions. This has led us to combine our hydrodynamic star cluster 
wind model with the stochastic dust injection, heating and cooling models to calculate the expected spectral energy 
distributions from super stellar clusters. 

We have followed the evolution of the grain size distribution, what changes drastically, the resultant 
spectrum. We have also considered the exit of dust grains as they stream out, coupled to the gas,
to compose the star cluster wind. For the latter, we have used a finite difference method.   

In this scenario, a certain mass of silicate and graphite dust, and an initial grain size distribution 
is injected into the intracluster medium. On top of this, the stellar winds are steady 
but the rate of supernova makes the dust injection an stochastic process. Therefore, dust is injected 
into the medium stochastically, and then heated and eroded before the next injection episode.

Several models were defined in order to quantify all these effects in the resultant infrared spectrum. 
Models which give more weight in the dust size distribution to small grains (as when dust injection is 
taking place), as well as models with larger values of the star cluster mechanical luminosity
(in which the SN rate increases leading to more persistent dust reservoirs), lead to an 
enhanced dust emission. The opposite situation occurs with more extended star clusters and 
larger adiabatic terminal speeds, which lead to a decrease in their dust emission. 
When dust injection ceases, the resultant SEDs change drastically and the
emission at NIR-MIR wavelengths vanishes due to thermal sputtering acting more effectively on small 
grains.

Despite the fact that our models imply the presence from hundredths to tenths of solar masses of dust 
inside the star cluster volume and transient strong NIR-MIR infrared excesses, the predicted 
SEDs are strong enough to be considered in order to explain the infrared excesses observed in bright 
young clusters and other stellar systems. In those cases, the combined action of many 
nearby star clusters, as well as higher SN rates in more massive clusters, could led to persistent 
infrared excesses. This and a detailed comparison with the observations of starburst galaxies will be 
addressed in a forthcoming communication.

\section{Acknowledgements}

We thank our anonymous referee for a detailed report full of valuable comments and helpful 
suggestions which greatly improved the paper. This study has been supported by CONACYT - M\'exico, 
research grants 167169, and 131913 and by the Spanish Ministry of Science and Innovation under 
the collaboration ESTALLIDOS (grants AYA2010-21887-C04-04 estallidosIV and AYA2013-47742-C4-2-P estallidos5). 
GTT also acknowledges the C\'atedra Severo Ochoa at the Instituto de Astrof\'isica de Canarias 
(IAC, Tenerife Spain) and the CONACYT grant 232876 for a sabbatical leave. SMG acknowledges 
Prof. B. T. Draine for his comments which helped to understand the normalization 
of the dust size distributions.

\appendix

\section{Appendix}
\label{app}

\subsection{The Cooling Function via Gas-Grain Collisions}
\label{app:1}

Following \citet{Dwek1987}, and keeping most of his notations and definitions, we calculated the cooling rate 
due to the gas-grain collisions in a dusty plasma with a normal chemical composition (one He atom per every 
ten H atoms):
\begin{eqnarray}
\label{eq:A5}
\Lambda_d = \frac{n_d}{n_e n} H_{coll} = 
\frac{1.4 m_{H} Z_d}{\rho_{d}} \left(\frac{32}{\pi m_{e}}\right)^{1/2} \pi (k_{B} T)^{3/2}  
\left[h_{e} + \frac{11}{23} \left( \frac{m_{e}}{m_{H}} \right)^{1/2}h_{n}   \right] ,
\end{eqnarray}
where $n$, $n_d$ and $n_e$ are the gas, dust and electron number density, $H_{coll}$ is the heating rate 
of a single grain due to collisions with incident gas particles and $k_{B}$ is the Boltzmann constant. 
Functions $h_{e}$ and $h_{H}$ are the effective grain heating efficiencies due to impinging electrons 
and nuclei, respectively:
\begin{eqnarray}
\label{eq:A6}
h_{e} = \int_{a_{min}}^{a_{max}}  \int^{\infty}_0 \frac{\zeta(a,E)}{2} x_{e}^2 e^{\displaystyle -x_{e}} a^{2} \derp{n_{i}}{a} \mbox{ d}x_{e} \mbox{ d}a  ,
\end{eqnarray}
\begin{eqnarray}
\label{eq:A7}
h_{n} = \int_{a_{min}}^{a_{max}} \left\{ 
          \left[1-\left(1+\frac{x_{H}}{2}\right) e^{\displaystyle -x_{H}}\right] 
      +   \frac{1}{2}\left[1-\left(1+\frac{x_{He}}{2}\right) e^{\displaystyle -x_{He}}\right] \right\} a^{2} \derp{n_{i}}{a} \mbox{ d}a ,  
\end{eqnarray}

where $\rho_{d} =4/3\pi \rho_{gr} \displaystyle \int_{a_{min}}^{a_{max}} a^{3} \derp{n_{i}}{a} \mbox{ d}a $ is the 
size-averaged dust density, $\rho_{gr}$ is the grain density, $x_{e}=E/k_{B}T$, $E$ is the energy of the impinging 
electron, $\zeta(a,E)$ is the fraction of the electron kinetic energy transfered to the dust grain, $x_{H}=E_{H}/k_{B}T$, $x_{He}=E_{He}/k_{B}T$ and the energies from the incident 
hydrogen and helium nuclei are $E_{H} = 133a$ keV, $E_{He} = 222a$ keV, where $a$ is measured in microns,

\begin{eqnarray}
\label{eq:A8}
\zeta(a,E) = 
\begin{cases} 
 \displaystyle 0.875, & \text{  if } E\leq E^* \\
              1-E_{f}/E,              & \text{otherwise} ,
\end{cases}          
\end{eqnarray}
      
where $E_{f}=max\{E',0.125E\}$, with $E^*$ and $E'$, the critical energy at which an electron penetrates the dust grain 
and the final energy of the electron after penetrating the dust grain, respectively. $E^*$ and $E'$ are obtained by solving 
the following system of non-linear equations based on experimental data
           
      \begin{eqnarray}
      \label{eq4}
      \hspace{-1.1cm}      
      \log{R(E^*)}    &=& \log{(4 a \rho_{gr}/3)}  = 0.146 \log{E^*}^2 + 0.5 \log{E^*} - 8.15, \\
      \log{R(E)}      &=&                           0.146 \log{E}^2   + 0.5 \log{E}   - 8.15,  \\
      \log{R(E')}     &=& \log{R(E)-R(E^*)}      = 0.146 \log{E'}^2  + 0.5 \log{E'}  - 8.15.
      \end{eqnarray}
     
\begin{figure}[htbp]
 \epsscale{1.0}
 \plotone{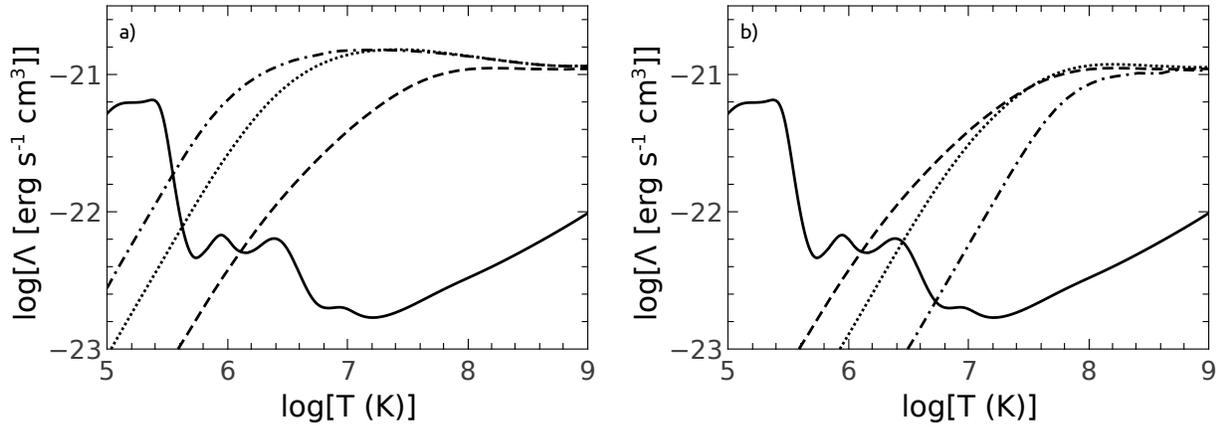}
 \caption[Cooling function for different \citetalias{MRN1977} dust size distributions as a function of temperature.]
 {Cooling function for different \citetalias{MRN1977} dust size distributions as a function of temperature. In panel a), 
 $a_{min}$ is set to  $0.001$ $\mu$m and $a_{max}$ takes the values $0.001$, $0.01$ and $0.5$ $\mu$m 
 (dashed, dotted, and dash-dotted lines, respectively).  In panel b), $a_{max}$ is set to  $0.5$ $\mu$m 
 and $a_{min}$ takes the values $0.001$, $0.01$ and $0.5$ $\mu$m (dashed, dotted,  and dash-dotted lines, 
 respectively). In these calculations, we assumed a dust-to-gas mass ratio of $Z_{d}=10^{-3}$. In both panels, 
 the interstellar cooling law for solar metallicity is presented as a solid curve.}
 \label{fig:DC_SD}
\end{figure}

In Figure \ref{fig:DC_SD}, we show examples of the dust cooling curves for different cases of the 
\citetalias{MRN1977} dust size distributions in which the value of the dust-to-gas mass ratio is set 
to $Z_{d}=10^{-3}$. Finally, we provide tables which contain the results of the calculations of the dust 
cooling function for different \citetalias{MRN1977} dust size distributions as a function of the gas temperature and 
normalized to the dust-to-gas mass ratio, for $a_{min}$ set to $0.001$ $\mu$m and different $a_{max}$ 
values (Table \ref{tab:3}), and for $a_{max}$ set to $0.5$ $\mu$m and different $a_{min}$ values (Table \ref{tab:4}).


\begin{table}[htp]
\centering
\caption{The Cooling Function with different \citetalias{MRN1977} dust size distribution 
(with $a_{min}= 0.001$ $\mu$m in all cases) normalized to the 
dust-to-gas mass ratio.}
\tiny
\label{tab:3}
\begin{tabular}{c c c c c c c c c }
\hline\hline
 $a_{max}$ ($\mu$m)               & 0.001        &    0.002     &      0.005   &   0.01       &      0.05    &     0.1      &  0.5     \\ \hline\hline
 T (K)&        \multicolumn{7}{c}{$\Lambda_{d}/Z_{d}$ (erg s$^{-1}$ cm$^3$)} \\ \cline{2-8}  &       \\ 
1.00E+04  &   8.787E-22  &   6.213E-22  &   3.930E-22  &   2.778E-22  &   1.234E-22  &   8.606E-23  &   3.565E-23   \\
1.25E+04  &   1.256E-21  &   8.884E-22  &   5.619E-22  &   3.973E-22  &   1.774E-22  &   1.252E-22  &   5.702E-23   \\
1.57E+04  &   1.763E-21  &   1.246E-21  &   7.883E-22  &   5.574E-22  &   2.490E-22  &   1.756E-22  &   8.000E-23   \\
1.97E+04  &   2.473E-21  &   1.749E-21  &   1.106E-21  &   7.821E-22  &   3.493E-22  &   2.464E-22  &   1.122E-22   \\
2.47E+04  &   3.470E-21  &   2.453E-21  &   1.552E-21  &   1.097E-21  &   4.900E-22  &   3.457E-22  &   1.575E-22   \\
3.09E+04  &   4.868E-21  &   3.442E-21  &   2.177E-21  &   1.539E-21  &   6.875E-22  &   4.850E-22  &   2.209E-22   \\
3.87E+04  &   6.830E-21  &   4.830E-21  &   3.054E-21  &   2.160E-21  &   9.646E-22  &   6.805E-22  &   3.100E-22   \\
4.86E+04  &   9.583E-21  &   6.776E-21  &   4.286E-21  &   3.030E-21  &   1.353E-21  &   9.547E-22  &   4.349E-22   \\
6.09E+04  &   1.344E-20  &   9.507E-21  &   6.013E-21  &   4.251E-21  &   1.899E-21  &   1.339E-21  &   6.102E-22   \\
7.63E+04  &   1.886E-20  &   1.334E-20  &   8.436E-21  &   5.965E-21  &   2.664E-21  &   1.879E-21  &   8.561E-22   \\
9.56E+04  &   2.646E-20  &   1.871E-20  &   1.184E-20  &   8.369E-21  &   3.737E-21  &   2.637E-21  &   1.201E-21   \\
1.20E+05  &   3.713E-20  &   2.625E-20  &   1.660E-20  &   1.174E-20  &   5.244E-21  &   3.699E-21  &   1.685E-21   \\
1.50E+05  &   5.209E-20  &   3.683E-20  &   2.330E-20  &   1.647E-20  &   7.357E-21  &   5.190E-21  &   2.364E-21   \\
1.88E+05  &   7.308E-20  &   5.168E-20  &   3.269E-20  &   2.311E-20  &   1.032E-20  &   7.282E-21  &   3.317E-21   \\
2.36E+05  &   1.025E-19  &   7.250E-20  &   4.586E-20  &   3.242E-20  &   1.448E-20  &   1.022E-20  &   4.653E-21   \\
2.96E+05  &   1.434E-19  &   1.017E-19  &   6.431E-20  &   4.548E-20  &   2.031E-20  &   1.433E-20  &   6.524E-21   \\
3.70E+05  &   1.995E-19  &   1.422E-19  &   9.010E-20  &   6.373E-20  &   2.847E-20  &   2.008E-20  &   9.132E-21   \\
4.64E+05  &   2.737E-19  &   1.978E-19  &   1.258E-19  &   8.907E-20  &   3.981E-20  &   2.807E-20  &   1.274E-20   \\
5.82E+05  &   3.672E-19  &   2.716E-19  &   1.744E-19  &   1.238E-19  &   5.540E-20  &   3.907E-20  &   1.767E-20   \\
7.29E+05  &   4.773E-19  &   3.651E-19  &   2.389E-19  &   1.703E-19  &   7.650E-20  &   5.395E-20  &   2.428E-20   \\
9.14E+05  &   5.998E-19  &   4.772E-19  &   3.213E-19  &   2.310E-19  &   1.045E-19  &   7.370E-20  &   3.300E-20   \\
1.15E+06  &   7.272E-19  &   6.032E-19  &   4.223E-19  &   3.077E-19  &   1.406E-19  &   9.932E-20  &   4.426E-20   \\
1.44E+06  &   8.536E-19  &   7.363E-19  &   5.399E-19  &   4.012E-19  &   1.863E-19  &   1.319E-19  &   5.851E-20   \\
1.80E+06  &   9.735E-19  &   8.691E-19  &   6.701E-19  &   5.109E-19  &   2.428E-19  &   1.724E-19  &   7.629E-20   \\
2.25E+06  &   1.080E-18  &   9.951E-19  &   8.063E-19  &   6.342E-19  &   3.111E-19  &   2.218E-19  &   9.808E-20   \\
2.82E+06  &   1.176E-18  &   1.110E-18  &   9.415E-19  &   7.667E-19  &   3.918E-19  &   2.810E-19  &   1.244E-19   \\
3.54E+06  &   1.264E-18  &   1.209E-18  &   1.068E-18  &   9.016E-19  &   4.850E-19  &   3.508E-19  &   1.561E-19   \\
4.44E+06  &   1.329E-18  &   1.296E-18  &   1.183E-18  &   1.033E-18  &   5.902E-19  &   4.316E-19  &   1.933E-19   \\
5.56E+06  &   1.383E-18  &   1.365E-18  &   1.280E-18  &   1.153E-18  &   7.051E-19  &   5.232E-19  &   2.371E-19   \\
6.97E+06  &   1.447E-18  &   1.419E-18  &   1.360E-18  &   1.258E-18  &   8.264E-19  &   6.248E-19  &   2.879E-19   \\
8.73E+06  &   1.485E-18  &   1.469E-18  &   1.425E-18  &   1.346E-18  &   9.494E-19  &   7.345E-19  &   3.459E-19   \\
1.09E+07  &   1.498E-18  &   1.508E-18  &   1.473E-18  &   1.415E-18  &   1.067E-18  &   8.484E-19  &   4.117E-19   \\
1.37E+07  &   1.506E-18  &   1.522E-18  &   1.505E-18  &   1.465E-18  &   1.174E-18  &   9.614E-19  &   4.849E-19   \\
1.72E+07  &   1.508E-18  &   1.520E-18  &   1.523E-18  &   1.497E-18  &   1.265E-18  &   1.068E-18  &   5.646E-19   \\
2.15E+07  &   1.505E-18  &   1.514E-18  &   1.532E-18  &   1.516E-18  &   1.336E-18  &   1.161E-18  &   6.494E-19   \\
2.70E+07  &   1.497E-18  &   1.503E-18  &   1.524E-18  &   1.518E-18  &   1.387E-18  &   1.238E-18  &   7.367E-19   \\
3.38E+07  &   1.483E-18  &   1.488E-18  &   1.502E-18  &   1.514E-18  &   1.419E-18  &   1.297E-18  &   8.224E-19   \\
4.24E+07  &   1.464E-18  &   1.468E-18  &   1.478E-18  &   1.496E-18  &   1.434E-18  &   1.336E-18  &   9.022E-19   \\
5.31E+07  &   1.441E-18  &   1.444E-18  &   1.451E-18  &   1.463E-18  &   1.434E-18  &   1.357E-18  &   9.716E-19   \\
6.66E+07  &   1.414E-18  &   1.416E-18  &   1.421E-18  &   1.429E-18  &   1.422E-18  &   1.363E-18  &   1.027E-18   \\
8.35E+07  &   1.384E-18  &   1.385E-18  &   1.389E-18  &   1.394E-18  &   1.400E-18  &   1.356E-18  &   1.068E-18   \\
1.05E+08  &   1.352E-18  &   1.353E-18  &   1.355E-18  &   1.359E-18  &   1.375E-18  &   1.340E-18  &   1.095E-18   \\
1.31E+08  &   1.319E-18  &   1.319E-18  &   1.321E-18  &   1.324E-18  &   1.339E-18  &   1.315E-18  &   1.109E-18   \\
1.64E+08  &   1.286E-18  &   1.286E-18  &   1.287E-18  &   1.289E-18  &   1.295E-18  &   1.290E-18  &   1.113E-18   \\
2.06E+08  &   1.253E-18  &   1.254E-18  &   1.254E-18  &   1.256E-18  &   1.258E-18  &   1.257E-18  &   1.110E-18   \\
2.58E+08  &   1.224E-18  &   1.224E-18  &   1.224E-18  &   1.225E-18  &   1.225E-18  &   1.219E-18  &   1.106E-18   \\
3.23E+08  &   1.197E-18  &   1.197E-18  &   1.198E-18  &   1.198E-18  &   1.197E-18  &   1.190E-18  &   1.097E-18   \\
4.05E+08  &   1.176E-18  &   1.176E-18  &   1.176E-18  &   1.176E-18  &   1.175E-18  &   1.167E-18  &   1.095E-18   \\
5.08E+08  &   1.160E-18  &   1.160E-18  &   1.160E-18  &   1.160E-18  &   1.159E-18  &   1.153E-18  &   1.094E-18   \\
6.37E+08  &   1.151E-18  &   1.152E-18  &   1.152E-18  &   1.152E-18  &   1.151E-18  &   1.145E-18  &   1.090E-18   \\
7.98E+08  &   1.151E-18  &   1.151E-18  &   1.151E-18  &   1.152E-18  &   1.151E-18  &   1.147E-18  &   1.099E-18   \\

\hline\hline
\end{tabular}
\end{table}

\begin{table}[htp]
\centering
\tiny
\caption{The Cooling Function with different \citetalias{MRN1977} dust size distribution (with $a_{max}=0.5$ $\mu$m in all cases) normalized to 
the dust-to-gas mass ratio.}
\label{tab:4}
\begin{tabular}{c c c c c c c c c }
\hline\hline
$a_{min}$ ($\mu$m)  & 0.001        &    0.002     &      0.005   &   0.01       &      0.05    &     0.1      &  0.5     \\ \hline\hline
 T (K)&        \multicolumn{7}{c}{$\Lambda_{d}/Z_{d}$ (erg s$^{-1}$ cm$^3$)} \\ \cline{2-8}  &       \\ 
1.00E+04  &   3.565E-23  &   2.625E-23  &   1.721E-23  &   1.234E-23  &   5.556E-24  &   3.930E-24  &   1.757E-24   \\
1.25E+04  &   5.702E-23  &   3.959E-23  &   2.504E-23  &   1.774E-23  &   7.946E-24  &   5.619E-24  &   2.513E-24   \\
1.57E+04  &   8.000E-23  &   5.555E-23  &   3.513E-23  &   2.489E-23  &   1.115E-23  &   7.883E-24  &   3.525E-24   \\
1.97E+04  &   1.122E-22  &   7.794E-23  &   4.928E-23  &   3.493E-23  &   1.564E-23  &   1.106E-23  &   4.946E-24   \\
2.47E+04  &   1.575E-22  &   1.093E-22  &   6.914E-23  &   4.900E-23  &   2.195E-23  &   1.552E-23  &   6.940E-24   \\
3.09E+04  &   2.209E-22  &   1.534E-22  &   9.701E-23  &   6.875E-23  &   3.079E-23  &   2.177E-23  &   9.736E-24   \\
3.87E+04  &   3.100E-22  &   2.152E-22  &   1.361E-22  &   9.646E-23  &   4.320E-23  &   3.055E-23  &   1.366E-23   \\
4.86E+04  &   4.349E-22  &   3.020E-22  &   1.910E-22  &   1.353E-22  &   6.061E-23  &   4.286E-23  &   1.917E-23   \\
6.09E+04  &   6.102E-22  &   4.237E-22  &   2.679E-22  &   1.899E-22  &   8.503E-23  &   6.012E-23  &   2.689E-23   \\
7.63E+04  &   8.561E-22  &   5.944E-22  &   3.759E-22  &   2.664E-22  &   1.193E-22  &   8.436E-23  &   3.773E-23   \\
9.56E+04  &   1.201E-21  &   8.340E-22  &   5.273E-22  &   3.738E-22  &   1.674E-22  &   1.183E-22  &   5.293E-23   \\
1.20E+05  &   1.685E-21  &   1.170E-21  &   7.399E-22  &   5.244E-22  &   2.348E-22  &   1.661E-22  &   7.426E-23   \\
1.50E+05  &   2.364E-21  &   1.642E-21  &   1.038E-21  &   7.357E-22  &   3.295E-22  &   2.330E-22  &   1.042E-22   \\
1.88E+05  &   3.317E-21  &   2.303E-21  &   1.456E-21  &   1.032E-21  &   4.622E-22  &   3.268E-22  &   1.462E-22   \\
2.36E+05  &   4.653E-21  &   3.231E-21  &   2.043E-21  &   1.448E-21  &   6.485E-22  &   4.586E-22  &   2.051E-22   \\
2.96E+05  &   6.524E-21  &   4.533E-21  &   2.867E-21  &   2.032E-21  &   9.099E-22  &   6.434E-22  &   2.877E-22   \\
3.70E+05  &   9.132E-21  &   6.360E-21  &   4.022E-21  &   2.851E-21  &   1.276E-21  &   9.027E-22  &   4.037E-22   \\
4.64E+05  &   1.274E-20  &   8.922E-21  &   5.643E-21  &   3.999E-21  &   1.791E-21  &   1.266E-21  &   5.664E-22   \\
5.82E+05  &   1.767E-20  &   1.251E-20  &   7.917E-21  &   5.611E-21  &   2.513E-21  &   1.777E-21  &   7.946E-22   \\
7.29E+05  &   2.428E-20  &   1.751E-20  &   1.111E-20  &   7.872E-21  &   3.525E-21  &   2.493E-21  &   1.115E-21   \\
9.14E+05  &   3.300E-20  &   2.441E-20  &   1.558E-20  &   1.105E-20  &   4.946E-21  &   3.497E-21  &   1.564E-21   \\
1.15E+06  &   4.426E-20  &   3.380E-20  &   2.185E-20  &   1.550E-20  &   6.940E-21  &   4.907E-21  &   2.195E-21   \\
1.44E+06  &   5.851E-20  &   4.632E-20  &   3.060E-20  &   2.174E-20  &   9.736E-21  &   6.885E-21  &   3.079E-21   \\
1.80E+06  &   7.629E-20  &   6.266E-20  &   4.270E-20  &   3.049E-20  &   1.366E-20  &   9.659E-21  &   4.320E-21   \\
2.25E+06  &   9.808E-20  &   8.350E-20  &   5.918E-20  &   4.271E-20  &   1.916E-20  &   1.355E-20  &   6.061E-21   \\
2.82E+06  &   1.244E-19  &   1.095E-19  &   8.113E-20  &   5.963E-20  &   2.689E-20  &   1.901E-20  &   8.503E-21   \\
3.54E+06  &   1.561E-19  &   1.413E-19  &   1.097E-19  &   8.271E-20  &   3.772E-20  &   2.668E-20  &   1.193E-20   \\
4.44E+06  &   1.933E-19  &   1.796E-19  &   1.458E-19  &   1.135E-19  &   5.292E-20  &   3.743E-20  &   1.674E-20   \\
5.56E+06  &   2.371E-19  &   2.251E-19  &   1.903E-19  &   1.534E-19  &   7.420E-20  &   5.251E-20  &   2.348E-20   \\
6.97E+06  &   2.879E-19  &   2.782E-19  &   2.439E-19  &   2.036E-19  &   1.038E-19  &   7.366E-20  &   3.295E-20   \\
8.73E+06  &   3.459E-19  &   3.394E-19  &   3.071E-19  &   2.649E-19  &   1.444E-19  &   1.032E-19  &   4.622E-20   \\
1.09E+07  &   4.117E-19  &   4.088E-19  &   3.799E-19  &   3.376E-19  &   1.988E-19  &   1.443E-19  &   6.485E-20   \\
1.37E+07  &   4.849E-19  &   4.865E-19  &   4.621E-19  &   4.214E-19  &   2.691E-19  &   2.001E-19  &   9.098E-20   \\
1.72E+07  &   5.646E-19  &   5.716E-19  &   5.527E-19  &   5.152E-19  &   3.561E-19  &   2.736E-19  &   1.276E-19   \\
2.15E+07  &   6.494E-19  &   6.621E-19  &   6.495E-19  &   6.166E-19  &   4.583E-19  &   3.659E-19  &   1.783E-19   \\
2.70E+07  &   7.367E-19  &   7.551E-19  &   7.499E-19  &   7.223E-19  &   5.719E-19  &   4.746E-19  &   2.470E-19   \\
3.38E+07  &   8.224E-19  &   8.465E-19  &   8.491E-19  &   8.266E-19  &   6.905E-19  &   5.940E-19  &   3.354E-19   \\
4.24E+07  &   9.022E-19  &   9.315E-19  &   9.413E-19  &   9.247E-19  &   8.060E-19  &   7.157E-19  &   4.413E-19   \\
5.31E+07  &   9.716E-19  &   1.005E-18  &   1.021E-18  &   1.011E-18  &   9.110E-19  &   8.300E-19  &   5.572E-19   \\
6.66E+07  &   1.027E-18  &   1.064E-18  &   1.086E-18  &   1.081E-18  &   9.989E-19  &   9.287E-19  &   6.724E-19   \\
8.35E+07  &   1.068E-18  &   1.107E-18  &   1.132E-18  &   1.132E-18  &   1.067E-18  &   1.008E-18  &   7.781E-19   \\
1.05E+08  &   1.095E-18  &   1.135E-18  &   1.162E-18  &   1.165E-18  &   1.113E-18  &   1.065E-18  &   8.625E-19   \\
1.31E+08  &   1.109E-18  &   1.149E-18  &   1.177E-18  &   1.182E-18  &   1.144E-18  &   1.103E-18  &   9.281E-19   \\
1.64E+08  &   1.113E-18  &   1.152E-18  &   1.180E-18  &   1.186E-18  &   1.161E-18  &   1.122E-18  &   9.670E-19   \\
2.06E+08  &   1.110E-18  &   1.147E-18  &   1.174E-18  &   1.180E-18  &   1.164E-18  &   1.133E-18  &   9.946E-19   \\
2.58E+08  &   1.106E-18  &   1.140E-18  &   1.166E-18  &   1.173E-18  &   1.163E-18  &   1.144E-18  &   1.023E-18   \\
3.23E+08  &   1.097E-18  &   1.129E-18  &   1.152E-18  &   1.158E-18  &   1.151E-18  &   1.139E-18  &   1.018E-18   \\
4.05E+08  &   1.095E-18  &   1.124E-18  &   1.144E-18  &   1.150E-18  &   1.147E-18  &   1.140E-18  &   1.023E-18   \\
5.08E+08  &   1.094E-18  &   1.118E-18  &   1.137E-18  &   1.142E-18  &   1.139E-18  &   1.134E-18  &   1.074E-18   \\
6.37E+08  &   1.090E-18  &   1.110E-18  &   1.125E-18  &   1.129E-18  &   1.124E-18  &   1.118E-18  &   1.080E-18   \\
7.98E+08  &   1.099E-18  &   1.116E-18  &   1.127E-18  &   1.129E-18  &   1.124E-18  &   1.118E-18  &   1.071E-18   \\
\hline\hline
\end{tabular}
\end{table}

\subsection{Stochastic Dust Temperature Distribution}
\label{app:2}

In order to calculate the temperature distribution of dust grains subject 
to a bath of free electrons in a hot gas, and thus the emission by such dust 
grains, we follow the schemes proposed by \citet{Dwek1986} and \citet{GuhathakurtaandDraine1989}
with a few extra considerations. In the Dwek's scenario, a dust 
grain with an initial temperature $T_{0}$, collides with a free electron with energy 
$E$ which transfers a fraction of its kinetic energy, $\zeta(a,E)$, to the dust 
particle. Depending on the size and chemical composition of the dust grain 
(because its heat capacity, $C(a,T_{d})$, is a function of both), the dust particle 
will be heated to a peak temperature $T_{peak}$, which is obtained from 
iteration of the equation

\begin{eqnarray}
\zeta(a,E) E = \int_{T_{0}}^{T_{peak}} C(a,T_{d}) \mbox{ d}T_{d} .
\end{eqnarray}

From $T_{peak}$, the dust particle starts to cool down and eventually, after many collisions, it acquires thermodynamic
equilibrium  unless the characteristic time for electron-grain collisions is larger than the 
grain cooling time, in which case the grain temperature will start to fluctuate \citep{Dwek1986,DwekandArendt1992}. 
The grain cooling time, $\tau_{cool}$, between $T_{peak}$ and some temperature $T_{d}$, is given by 

\begin{eqnarray}
\tau_{cool} = \int_{T_{d}}^{T_{peak}}
              \frac{C(a,T_{d})\mbox{ d} T_{d}}{|4\pi a^2 \sigma \langle Q_{abs} \rangle T_{d}^{4}|} ,  
\end{eqnarray}

where $\sigma$ is the Stefan-Boltzmann constant and $\langle Q_{abs} \rangle$ is the dust 
absorption efficiency, $Q_{abs}(\lambda,a)$, averaged by the Planck function, $B_{\lambda}(T_{d})$ 
(in terms of the wavelength, $\lambda$):

\begin{eqnarray}
\langle Q_{abs} \rangle = \frac{1}{\sigma T_{d}^4}\int_{0}^{\infty} \pi Q_{abs}(\lambda,a) 
                          B_{\lambda}(T_{d}) \mbox{ d}\lambda  .
\end{eqnarray}

The values of $C(a,T_{d})$ for silicate and graphite grains were taken from \citet{Dwek1986} and from \citet{DraineandAnderson1985}, 
while the values of $Q_{abs}(\lambda,a)$ were obtained from the data files provided in the DustEM code 
\footnote{http://www.ias.u-psud.fr/DUSTEM} \citep{Compiegneetal2011}. The characteristic 
time between successive electron collisions with a dust grain, $\tau_{coll}$, is calculated as
\citep{Bocchioetal2013}:

\begin{eqnarray}
\label{eq:taucoll}
 \tau_{coll}^{-1}= \pi a^2 n \sqrt{\frac{3 k_{B} T }{m_e}},
\end{eqnarray}

where $m_{e}$ is the mass of the electron. The fraction of time in which a dust grain 
can be found in the temperature interval $T_{d}+\mbox{ d}T_{d}$ after a collision with 
an electron is \citep{Purcell1976}:

\begin{eqnarray}
P(a,E,T_{d},T_{0}) \mbox{ d}T_{d} = 
\begin{cases} 
 \displaystyle \frac{C(a,T_{d})}{4\pi a^2 
          \sigma \langle Q_{abs} \rangle T_{d}^4}\frac{e^{-\tau_{cool}/\tau_{coll}}}{\tau_{coll}}, & \text{  if } T_{d}\leq T_{peak} \\
    \mbox{         } 0,              & \text{otherwise} .
\end{cases}          
\end{eqnarray}

One can obtain now the probability, $G(a,T_{d},T_{0})$, that a dust grain is to be found 
between $T_{d}$ and $T_{d}+\mbox{ d}T_{d}$ if one integrates the above quantity over all the 
electron energies according to the Maxwell-Boltzmann distribution

\begin{eqnarray}
\label{eq:Ge}
G(a,T_{d},T_{0}) = \pi a^2 n \tau_{coll} \int_{0}^{\infty} 
P(a,E,T_{d},T_{0}) f(E) v(E) \mbox{ d}E .
\end{eqnarray}

By evaluating equation \ref{eq:Ge}, we obtain the temperature distribution of a population of grains with the same initial 
temperature $T_{0}$, size and chemical composition. In the above equation, $f(E)$ and $v(E)$ are the Maxwell-Boltzmann 
distribution of energy and the speed of the impinging electron, respectively. In order to obtain the temperature distribution of grains with a wide 
range of initial temperatures, we will employ the stochastic matrix method described by 
\citet{GuhathakurtaandDraine1989} and \citet{Marengo2000}.

Let $\mathbf{A}_{T_{d}^{i},T_{0}^{j}}$, be an $N\times N$ stochastic matrix, which describes the probability (per unit time) of a grain to
make a transition between $T_{0}$ and some temperature $T_{d}$. The entries of $\mathbf{A}_{T_{d}^{i},T_{0}^{j}}$ are obtained from evaluation 
of equation \ref{eq:Ge}:

\begin{eqnarray}
      \label{eq:A}
 \mathbf{A}_{T_{d}^{i},T_{0}^{j}}        =
 \begin{pmatrix}
  G(a,T_{d}^{1},T_{0}^{1}) & G(a,T_{d}^{1},T_{0}^{2}) & \cdots & G(a,T_{d}^{1},T_{0}^{j}) \\
  G(a,T_{d}^{2},T_{0}^{1}) & G(a,T_{d}^{2},T_{0}^{2}) & \cdots & G(a,T_{d}^{2},T_{0}^{j}) \\
  \vdots  & \vdots  & \ddots & \vdots  \\
  G(a,T_{d}^{i},T_{0}^{1}) & G(a,T_{d}^{i},T_{0}^{2}) & \cdots & G(a,T_{d}^{i},T_{0}^{j}) 
 \end{pmatrix}.
\end{eqnarray}      

In our case we employed a logarithmic grid for $T_{d}$ and $T_{0}$, from $1$ K to $1100$ K and $N=125$.

Let now $G_{n=0}^{i}$ be an initial temperature distribution given by a column vector which comes from evaluating equation \ref{eq:Ge} with
$T_{0}=T_{0}^{trial}$, a trial initial temperature:

\begin{eqnarray}
      \label{eq:G0}
  G_{n=0}^{i}     =
 \begin{pmatrix}
  G(a,T_{d}^{1},T_{0}^{trial}) \\
  G(a,T_{d}^{2},T_{0}^{trial}) \\
           \vdots          \\
  G(a,T_{d}^{i},T_{0}^{trial}) 
 \end{pmatrix}.
\end{eqnarray}  

We apply the stochastic matrix to the initial temperature distribution to obtain a new stochastic temperature distribution,
$G_{n=1}^{i}= G_{n=0}^{i} \mathbf{A}_{T_{d}^{i},T_{0}^{trial}}$. We continue to iteratively apply the stochastic matrix,
 
\begin{eqnarray}
      \label{eq:G1}
 G_{n+1}^{i}      = G_{n}^{i} \mathbf{A}_{T_{d}^{i},T_{0}^{j}},
 \end{eqnarray}
 
until the condition $(\mathbf{I}- \mathbf{A}_{T_{d}^{i},T_{0}^{j}})G_{n+1}= 0$, with $\mathbf{I}$ the identity matrix, 
is fulfilled. This condition ensures that, after many discrete heating events, the temperature distribution does not change under the 
application of the stochastic matrix; this is the steady state temperature distribution, $G(a,T_{d})$. 

Big grains ($\gtrsim 0.1$ $\mu$m), with large cross sections and heat capacities, are more likely to reach thermodynamic equilibrium due 
to very frequent collisions. In that case, their temperature distribution approaches a delta function around the equilibrium temperature, $T_{eq}$, 
which can be obtained by equating the heating and cooling rates:

\begin{eqnarray}
\pi a^2  n \int_{0}^{\infty} f(E) v(E) \zeta(a,E) E \mbox{ d}E = 4\pi a^2 \sigma \langle Q_{abs} \rangle T_{eq}^4 .
\end{eqnarray}

Once the dust temperature distribution is known, the infrared flux can be calculated from equation 15 of \citet{DwekandArendt1992}.
\bibliographystyle{apj}
\bibliography{Infrared}

\end{document}